\documentstyle[prd,aps,epsfig,eqsecnum,amssymb,amsmath,floats]{revtex}
\def\abstract{\par
\ifpreprintsty %
\vskip2.5pc
\begin{center}%
{\large \abstractname\par}%
\end{center}%
\vskip.5pc
\fi
\bgroup
\ifdim\prevdepth=-1000pt \prevdepth0pt\fi
\hsize2\columnwidth
\if@twocolumn\else\leftskip=0.10753\textwidth \rightskip\leftskip\fi
\dimen0=-\prevdepth \advance\dimen0 by17.5pt \nointerlineskip
\small\vrule width 0pt height\dimen0 \relax
}

\begin{document}

\title{Astrophysical tau neutrinos and their detection by large neutrino telescopes}
\author{
Edgar Bugaev,$^1$ Teresa Montaruli,$^{2,3}$ and Igor Sokalski$^{3,1}$\\[1mm]
{\it $^1$ INR, 60th October Anniversary Prospect 7a, 117312, Moscow, Russia\\
$^2$ Physics Department, Bari University, Via Amendola 173, 70126 Bari, Italy\\      
$^3$ INFN/Bari, Via Amendola 173, 70126 Bari, Italy\\}
}
\date{
\today
}
\twocolumn[
\begin{@twocolumnfalse}

\maketitle
\begin{abstract}
We present results of the detailed Monte Carlo calculation of the rates of double-bang events
in 1 km$^3$ underwater neutrino telescope with taking into account the effects of $\tau$-neutrino
propagation through the Earth. As an input, the moderately optimistic theoretical predictions
for diffuse neutrino spectra of AGN jets are used.
\end{abstract}
\vskip12.5pt
\end{@twocolumnfalse}
]

\section{Introduction}
\label{sec:1}
During these years there have been many discussions \cite{1,2} on the 
possibilities of very high energy $\tau$-neutrino detection by large 
neutrino telescopes, which would prove oscillations of muon neutrinos
of astrophysical origin ("astrophysical long baseline
experiment" \cite{3}). 

It is known now that neutrino oscillations definitely exist. Nevertheless, the  detection of 
astrophysical $\tau$-neutrinos is still very interesting and useful, e.g., for a better 
understanding of neutrino properties as well as properties of astrophysical sources \cite{4}.

Intrinsic flux of $\tau$-neutrinos from a typical source in which there is acceleration of protons
and production of neutrinos through $pp$- and $p\gamma$-reactions is very small. The dominant 
channel of $\tau$-neutrino birth is through the inclusive production of charmed mesons \cite{5},
\begin{equation}
pp,p\gamma \longrightarrow D^{+}_{S} + X,\;\;\;\;\;\; D^{+}_{S} \longrightarrow \tau\nu_{\tau}\;\;,\nonumber 
\end{equation}
and the smallness of $\nu_{\tau}$-flux  follows from the relations 
\begin{equation}
\frac{\sigma_{\gamma p\to DX}}{\sigma_{\gamma p \to \pi X}} \alt 10^{-3}\;,\;\;
\frac{Br(D_{S}^{+}\to\nu_{\tau})}{Br(\pi\to\nu_{\mu})}\sim 7\cdot 10^{-2}.\nonumber
\end{equation}
Therefore, the production of $\nu_{\tau}$ at source is negligible, and
if all muons can decay, it can be expected for the 3 flavors
the following proportion: 
\begin{equation}
\Phi^{0}_{std}=\{\phi^0_{e},\phi^0_{\mu},\phi^0_{\tau}\}=\{\frac{1}{3},\frac{2}{3},0\}.\nonumber
\end{equation}

Assuming maximal atmospheric mixing and $U_{e\;3}=0$, that is:
\begin{equation}
\sin \theta_{1\;3}=0\;,\;\;\; \sin 2 \theta_{2\;3}=1\;\;\;\nonumber
\end{equation}
one obtains, after averaging over many oscillations,
the ideal equipartition between neutrino flavors in the 
astrophysical neutrino flux at Earth, 
\begin{equation}
\Phi_{std}=\{\frac{1}{3}, \frac{1}{3}, \frac{1}{3}\},\nonumber
\end{equation}
independently on the value of the solar mixing angle $\theta_{1\;2}$. 
Moreover, in the case when there is maximal atmospheric mixing and $U_{e \;3}=0$, for any 
proportions between flavors at the source one has, after propagation, the equality
$\phi_{\mu}=\phi_{\tau}$ \cite{4}.

The experimental verification of the relation $\phi_{\mu}=\phi_{\tau}$ by detection of
astrophysical diffuse fluxes of $\nu_{\tau}$ and $\nu_{\mu}$ with flavor 
identification capabilities
is very important because the inequality $\phi_{\mu}\ne\phi_{\tau}$ is predicted in some exotic
scenarios, e.g. \cite{4}: i) if CPT invariance is violated in the neutrino sector and
the neutrino and antineutrino mixing matrices do not coincide, $\phi_{\mu}$, in general, is
not equal to $\phi_{\tau}$,
ii) if neutrinos can decay \cite{6} {\it en route } from the sources to the Earth, the neutrino fluxes 
and flavor ratios are very sensitive to uncertanties in the neutrino mixing matrix and strongly 
depend on the hierarchy of neutrino masses. So, the inequality $\phi_{\mu}\ne\phi_{\tau}$ 
would reveal unconventional neutrino physics. 

Inside of some astrophysical and cosmological objects there are processes in which very energetic
quarks  and gluons fragmenting into jets of hadrons are expected to be produced. Even the usual 
inclusive spectrum of high energy $pp$- or $p\gamma$-collisions always contains the jet component
(described well by perturbative QCD). Jets can appear in decays of (hypothetical) supermassive 
particles which, in turn, are produced in decays of topological defects \cite{7}, in processes
of superheavy dark matter annihilation inside of stars \cite{8}, in a process of Hawking evaporation
of primordial black holes \cite{9}. It was shown in \cite{10} that decays of top quarks contained
in the  jets lead to production of $\tau$-neutrino flux from the jets, which is of the same order
as $\nu_{\mu},\nu_{e}$ fluxes (at large neutrino energies, close to the value of the jet mass).
This means that the intrinsic $\tau$-neutrino flux from the sources in which the jet phenomena are 
essential is, in general, not suppressed (in comparison with intrinsic $\nu_{\mu},\nu_{e}$ fluxes).

We studied in this paper the possibility of detection of extragalactic diffuse $\tau$-neutrino 
fluxes by large ($\sim1$km$^{3}$) underwater neutrino telescopes. In Sec.\ref{sec:2} we discuss
the general problems connected with the detection of very high energy $\tau$-neutrinos 
($E_{\nu}\agt 10^{6}\text{GeV}$). In Sec.\ref{sec:3} we present the diffuse neutrino spectra
for the AGN jets chosen for the numerical calculations. Details of the calculations, results and
conclusions are given in Sec.\ref{sec:4}.

\section{Detection of $\tau$-neutrinos}
\label{sec:2}
The main feature of charge current interactions of $\tau$-neutrinos is the fact that 
$\tau$-leptons decay long before they lose 
a large fraction of their energy. This leads 
to the absence of absorption of $\tau$-neutrinos during their propagation through the Earth
and , as a consequence, to the characteristic modification of the neutrino spectrum after
crossing the Earth (there is a pile-up of events with energies around $\sim 100\text{TeV}$,
if the incident neutrino spectrum is not too steep).  At neutrino energies $\sim 10^{6}-10^{7} \text{GeV}$ 
$\tau$-lepton track-length 
becomes larger than $\sim 100 \text{m}$ (in water),
it becomes possible to identify $\tau$-neutrino interactions in large neutrino telescopes
by selection of 
"double-bang" or "lollipop" events \cite{1}. 
A double-bang event
consists of two showers and a $\tau$-lepton track connecting them. The first (hadronic) shower is
initiated by the charge current interaction of $\tau$-neutrino and the second one (hadronic or 
electromagnetic) is produced by the $\tau$-lepton decay. In lollipop events one of these showers
is not detected (e.g., if the $\tau$-neutrino interaction takes place outside of the detector).

The  detection of events with double-bang structure would reliably indicate the appearance of
$\tau$-neutrinos in the detector because this signature is unique. However, the probability of
observing a double-bang event $P_{DB}$ decreases with neutrino energy for $E_{\nu} > 10^8\text{GeV}$
(due to 
$\tau$-lepton range increase).
The lollipop structure is not so spectacular
although the observation probability $P_L$ of lollipop events is at $E_{\nu}\sim 10^{8}\text{GeV}$
considerably (by a factor of 10  \cite{11}) larger than $P_{DB}$.

Diffuse astrophysical and cosmological neutrino fluxes in the ultra-high energy region, 
$E_{\nu}>10^9\text{GeV}$, (GZK neutrinos (see, e.g., \cite{12}), hypothetical neutrino fluxes
from topological defects \cite{13}, from Z-bursts \cite{14} etc.) are too small and inaccessible
for a study by underwater neutrino telescopes. Registration of $\tau$-neutrinos of such energies
can be accomplished by detecting the air showers caused by $\tau$-leptons produced by 
neutrino-nucleon interactions far away from the detector ("Earth-skimming idea" \cite{15}).
The huge modern air shower detectors (e.g., the HiRes \cite{16} or the Pierre Auger
detector \cite{17}) 
could be suitable for such experiments, but expected event rates are
low.

\section{Diffuse neutrino spectra from AGN jets}
\label{sec:3}

\begin{figure}[t!]
\label{fig:fig1}
\epsfig{file=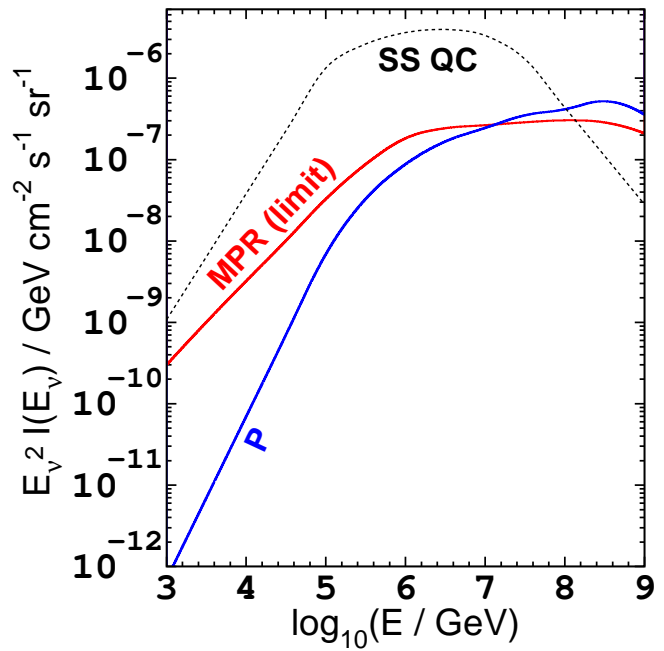,width=\columnwidth}
\caption{Different theoretical predictions for diffuse 
neutrino ($\nu_{\mu} + \tilde \nu_{\mu}$)
spectra from AGNs. MPR(limit): upper bound for $\nu_{\mu}$ 
flux from AGN jets \protect\cite{18}, P: $\nu_{\mu}$
flux for proton blazar model of \protect\cite{21}, 
SSQC: prediction of $\nu_{\mu}$ flux from radio-quiet AGNs \protect\cite{22}.}
\end{figure}

The concrete calculations were carried out with two theoretical diffuse neutrino spectra from AGNs.

1. The curve denoted by "MPR(limit)" in Fig.1 is the generic upper bound on the diffuse neutrino 
spectrum from AGN jets obtained by Mannheim, Protheroe, and Rachen \cite{18}. This limit was 
found using the assumption that 
the AGN source is not completely optically thin for a cosmic ray flux: due to opacity effects
there is a spectral break between $10^7\;\; GeV$ and $10^{11}\;\; GeV$ in the escaping cosmic ray 
spectrum (the resulting cosmic ray spectrum is a combination of several model spectra with 
different values  of the spectral break and with the fixed $E_{max}$ ($E_{max}=10^{11}\;\;GeV$),
and the normalization of the resulting spectrum is such that the cosmic ray intensity does 
not exceed  the proton spectrum estimated from observations). Neutrinos 
are mostly produced in decays of mesons  
from accelerated proton interactions with synchrotron 
photons emitted by high energy electrons in the jets. It is noted in \cite{18} 
that 
various models of
AGN jets (see, e.g., \cite{19,20,21}) predict diffuse neutrino spectra which are compatible with the MPR
limit (at the important neutrino energy range below $\sim 10^7\;\;GeV$).

2. The model by Protheroe \cite{21} (the curve denoted by "P" in Fig.1) is the optically thick proton 
blazar model, in which it is assumed that the target for the pion 
production is provided by the thermal UV photons emitted from the accretion disc rather than
by the synchrotron photons produced in the jets.
The diffuse neutrino spectrum is normalized to the experimental data on $\gamma$-ray background.
It is  seen from the figure that the prediction of
this model is, indeed, not in the contradiction with the MPR bound, if $E_{\nu}\le 10^{7} \;\;GeV$.
The proton blazar model, developed by Mannheim \cite{19} (in which accelerated protons interact
with the synchrotron radiation of jet electrons) predicts very similar diffuse neutrino spectrum.

One should note that the prediction given by Stecker and Salamon \cite{22} for diffuse neutrino spectrum
from radio-quiet AGNs is much more optimistic (SSQC curve in Fig.1) at neutrino energies
$E_{\nu}\sim 10^{6}-10^{7}\;\;GeV$. Their model, however, predicts a non-thermal spectrum of
X-rays at energies below $\sim 500\;\; KeV$(these X-rays result from reprocessing of gamma-rays from
pion decays by electromagnetic cascades due to the high photon density near the black hole) while
the observed X-ray spectra of radio-quiet AGNs (as well as the spectrum of the diffuse X-ray 
background) turn over steeply below $\sim 100\;\;KeV$ with no sign for a nonthermal component \cite{23}.
Therefore, the existing X-ray data cannot be used for a normalization of the radio-quiet AGN
neutrino spectrum (authors of \cite{22} normalized their spectrum, rather arbitrarily, on $30 \%$ 
of the diffuse X-ray background). This is the reason why we did not use SSQC flux in our calculations.
Besides, SSQC neutrino flux is now almost excluded by measurements on AMANDA \cite{24} and Baikal \cite{25}
neutrino telescopes.

\section{Results and discussions}
\label{sec:4}
First estimates of double-bang event rates in km$^3$ neutrino telescope were done in \cite{26},
for down-going $\tau$-neutrinos only. Detailed calculations of $\tau$-neutrino propagation
through the Earth carried out by Dutta et al. \cite{27} 
(using power law incident neutrino spectra) showed that, in spite of the absence of neutrino absorption, the contribution
of lower hemisphere 
events to the total event rate in a telescope cannot be too large.
This is due to a strong
decrease of neutrino energies during propagation. It was 
concluded in \cite{27} that the most promising
way to detect $\tau$-neutrino astrophysical fluxes is an analysis of shower events in the telescope.

In this work we present results on Monte Carlo calculations of rates of double-bang events
in a 1 km$^3$ telescope for both hemispheres using the diffuse neutrino spectra shown on Fig.1.
We simulate both DIS CC and NC neutrino interactions for all flavors in the energy range 
$10^3\text{GeV}<E_{\nu}<10^9\text{GeV}$ using the CTEQ3\_~DIS 
structure functions. For $\tau$-lepton
propagation we used the MUM code \cite{28} that was updated accounting for the corrections 
for photonuclear interactions of $\tau$-leptons \cite{29}. Some details of the calculations
and the resulting neutrino fluxes of all flavors after propagation through the Earth were presented
in our previous work \cite{30}.

The rate of totally contained double-bang events is given by the formula
\begin{eqnarray}
N=2\pi\rho N_A \int_{-1}^{1}\int_{E_{min}}^{\infty}  V_{eff}(E_{\nu_{\tau}},\theta)
I(E_{\nu_{\tau}}, \theta)\times\nonumber\\
\sigma^{CC}(E_{\nu_{\tau}}) d E_{\nu_{\tau}}d(\cos\theta).\nonumber
\end{eqnarray}
Here, $N_A$ is the Avogadro number, $\rho$ is the density of the medium (we use 
$\rho=1\text{g cm}^{-3}$ which is close to sea water/ice density), 
$I(E_{\nu_{\tau}}, \theta)$ is
the differential $\nu_{\tau}$ flux entering the detector at zenith angle $\theta$, $\sigma^{CC}$ is the total
deep inelastic neutrino cross section (charge current part).
$V_{eff}$ is the effective volume given by the relation:
\begin{eqnarray}
V_{eff}(E_{\nu_{\tau}},\theta)=S_{p}(\theta) (L-R_{\tau}(E_{\nu_{\tau}})),\nonumber
\end{eqnarray}
where $S_p$ is the projected area for tracks generated isotropically in azimuth at the fixed
$\theta$ directions on a detector of parallelepiped form, $L$ is the geometrical distance
between entry and exit point, $R_{\tau}$ is the $\tau$ lepton range. The minimum value of
the neutrino energy, $E_{min}=2\cdot 10^6\text{GeV}$,corresponds to a $\tau$-lepton range
of $R_{\tau}^{min}=70\text{m}$. 

The double-bang 
topology can be considered as a background free signature of $\tau$-neutrino 
appearance \cite{1} since there should be no atmospheric muons which could produce in the detector
2 showers with comparable amounts of photons. Nevertheless, this reasonable  assumption needs
to be verified through a full simulation.

Concrete calculations were done for two geometries of the neutrino detector:
\begin{eqnarray}
1\times 1\times 1\;\;\; \text{km}^3 \;\;\;\;\;\;\; (\text{IceCube-like \cite{31}}),\nonumber\\
1.4\times 1.4\times 0.6\;\;\;\text{km}^3\;\;\;\;\;\;\;(\text{NEMO-like \cite{32}}).\nonumber
\end{eqnarray}

In Table~1 the corresponding double-bang rates for lower, upper and both hemispheres are shown.
\begin{tabular}{c}
~~~~~~~~~~~~~~~~~~~~~~~~~~~~~~~~~~~~~~~~~~~~~~~~~   ~\\
~~~~~~~~~~~~~~~~~~~~~~~~~~~~~~~~~~~~~~~~~~~~~~~~~   ~\\
TABLE~1. ~~~~~~~~ Number of totally contained double-\\
bang events in km$^3$ detector per year.~~~~~~~~~~~~~\\
\\
\hline
\hline
~~Spectrum~~~~~~~  ~~~~Rate~$(N_{-2\pi} / N_{2\pi} / N_{4\pi})$ \\
\hline
~~~~~~~~~~~~~~~~~~~~~~~~~  ~~~~~~~~~~~~~~~NEMO-like~~~~~~~~~~~~~ \\
~~~~~~~\cite{21}~~~~~~~~~  ~~~~~~~~~~1.0~~~~~~2.1~~~~~~3.1~~~~~ \\
~~~~~~~\cite{18}~~~~~~~~~  ~~~~~~~~~~1.4~~~~~~3.1~~~~~~4.5~~~~~ \\
\hline
~~~~~~~~~~~~~~~~~~~~~~~~~  ~~~~~~~~~~~~~~IceCube-like~~~~~~~~~~ \\
~~~~~~~\cite{21}~~~~~~~~~  ~~~~~~~~~~0.7~~~~~~1.6~~~~~~2.3~~~~~ \\
~~~~~~~\cite{18}~~~~~~~~~  ~~~~~~~~~~1.0~~~~~~2.3~~~~~~3.3~~~~~ \\
\hline
~~~~~~~~~~~~~~~~~~~~~
\end{tabular}

Calculated values for upper hemisphere are $3-6$ times lower compared to \cite{26}
since more optimistic predictions for diffuse neutrino fluxes were used there.

Finally, we found using the theoretical predictions in \cite{18,21} for AGN diffuse neutrino fluxes
that in a 1~km$^3$ neutrino telescope one can expect a marginally observable rate of double-bang 
events from $\tau$-neutrinos.

\end{document}